\documentclass[sigconf]{acmart}

\usepackage{booktabs} % For formal tables
\usepackage[english]{babel}
\usepackage[export]{adjustbox}
\usepackage{mathrsfs}
\usepackage{csquotes}
\usepackage{multirow}
\usepackage{multicol}
\usepackage{array}
\usepackage{arydshln}
\usepackage{booktabs}
\usepackage{subfig}
\usepackage{siunitx}
\usepackage[inline]{enumitem}
\usepackage{xspace}
\usepackage{etoolbox}
\makeatletter
\patchcmd\@combinedblfloats{\box\@outputbox}{\unvbox\@outputbox}{}{%
   \errmessage{\noexpand\@combinedblfloats could not be patched}%
}%
 \makeatother
 
\newlist{inlinelist}{enumerate*}{1}
\setlist*[inlinelist,1]{%
  label=(\roman*),
}
\author{Mohammad Aliannejadi}
\affiliation{%
  \institution{Universit{\`a} della Svizzera italiana (USI)}
}
\email{mohammad.alian.nejadi@usi.ch}

\author{Hamed Zamani}
\affiliation{%
  \institution{University of Massachusetts Amherst}
}
\email{zamani@cs.umass.edu}

\author{Fabio Crestani}
\affiliation{%
  \institution{Universit{\`a} della Svizzera italiana (USI)}
}
\email{fabio.crestani@usi.ch}

\author{W. Bruce Croft}
\affiliation{%
  \institution{University of Massachusetts Amherst}
}
\email{croft@cs.umass.edu}

\newcommand{\partitle}[1]{\vspace{2mm}\noindent\textbf{#1}}

\newcommand{\unimobile}{UniMobile}

\newcommand{\ntar}{NTAS}

\begin{document}

\title[Target Apps Selection: Towards a Unified Search Framework for Mobile Devices]{Target Apps Selection:\\Towards a Unified Search Framework for Mobile Devices}

\begin{abstract}

With the recent growth of conversational systems and intelligent assistants such as Apple Siri and Google Assistant, mobile devices are becoming even more pervasive in our lives. As a consequence, users are getting engaged with the mobile apps and frequently search for an information need in their apps. However, users cannot search within their apps through their intelligent assistants. This requires a \emph{unified mobile search} framework that identifies the target app(s) for the user's query, submits the query to the app(s), and presents the results to the user. In this paper, we take the first step forward towards developing unified mobile search. In more detail, we introduce and study the task of \emph{target apps selection}, which has various potential real-world applications.
To this aim, we analyze attributes of search queries as well as user behaviors, while searching with different mobile apps. The analyses are done based on thousands of queries that we collected through crowdsourcing. We finally study the performance of state-of-the-art retrieval models for this task and propose two simple yet effective neural models that significantly outperform the baselines. Our neural approaches are based on learning high-dimensional representations for mobile apps. Our analyses and experiments suggest specific future directions in this research area.

\end{abstract}
% \keywords{mobile information retrieval;  mobile usage understanding; query analysis; neural networks}

\copyrightyear{2018} 
\acmYear{2018} 
\setcopyright{acmcopyright}
\acmConference[SIGIR '18]{The 41st International ACM SIGIR Conference on Research and Development in Information Retrieval}{July 8--12, 2018}{Ann Arbor, MI, USA}
\acmBooktitle{SIGIR '18: The 41st International ACM SIGIR Conference on Research and Development in Information Retrieval, July 8--12, 2018, Ann Arbor, MI, USA}
\acmPrice{15.00}
\acmDOI{10.1145/3209978.3210039}
\acmISBN{978-1-4503-5657-2/18/07}

\maketitle

\section{Introduction}
\label{sec:intro}

Recent years have witnessed a rapid growth in the use of mobile devices, enabling people to access the Internet in various contexts. More than 77\% of Americans now own a smartphone\footnote{\url{http://www.pewinternet.org/fact-sheet/mobile/}}, with an increasing trend in terms of the time people spend on their phones. As of 2016, the average U.S. user spends 5 hours on mobile devices per day, with just 8\% of it spent in the phone's browser. In fact, people spend most of their time (72\%) using apps that have their own search feature\footnote{\url{http://flurrymobile.tumblr.com/post/157921590345/us-consumers-time-spent-on-mobile-crosses-5}}. Moreover, Google Play Store now features more than 3.5 million apps and users install an average of 35 mobile apps on their phones, using half of them regularly\footnote{\url{https://www.thinkwithgoogle.com/advertising-channels/apps/app-marketing-trends-mobile-landscape/}}. 

More recently, with the release of intelligent assistants, such as Google Assistant and Apple Siri, people are experiencing mobile search through a single voice-based interface. These systems introduce several research challenges. Given that people spend most of their times in apps and, as a consequence, most of their search interactions would be with apps (rather than a browser), one limitation is that users are unable to use a conversational system to search within many apps.
This suggests the need for a \emph{unified search framework} that \textit{replaces all the search boxes in the apps, with a single search box}. With such a framework, the user can submit a query through this system which will identify the target app(s) for the issued query. The query is then routed to the identified target apps and the results are displayed in a unified interface.

In this work, we are particularly interested in taking the first step towards developing a unified search framework for mobile devices by introducing and studying the task of \textit{target apps selection}, which is defined as identifying the target app(s) for a given query. To this end, we built a collection of cross-app search queries through crowdsourcing, which is released for research purposes\footnote{Available at \url{http://aliannejadi.github.io/unimobile.html}}. Our crowdsourcing experiment consists of two parts: we initially asked crowdworkers to explain their latest search experience on their smartphones and used them to define various realistic mobile search tasks. Then, we asked another set of workers to select the apps they would choose to complete the tasks as well as the query they would submit. We investigate various aspects of user behaviors while completing a search task. For instance, we show that users choose to complete most of the search tasks using two apps. In addition, we demonstrate that for the majority of the search tasks, most of the users prefer \textit{not} to use Google Search.

From the lessons learned from our data analysis, we propose two simple yet efficient neural target apps selection models. Our first model looks at the problem as a ranking task and produces a score for a given query-app pair. We study two different training settings for this model. Our second framework, on the other hand, casts the problem as a multi-label classification task. Both neural approaches, called NTAS, learn a high-dimensional representation for each app. Our experiments demonstrate that our model significantly outperforms a set of state-of-the-art models in this task.

In summary, the main contributions of this paper include:
\begin{itemize}[leftmargin=*]
    \item Designing and conducting two crowdsourcing tasks for collecting cross-app search queries for real-life search tasks. The tasks and queries are publicly available for research purposes.
    \item Presenting the first study of user behaviors while searching with different apps as well as their search queries. In particular, we study the attributes of the search queries that are submitted to different apps and user behaviors in terms of the apps they chose to complete a search task.
    \item Proposing two neural models for target apps selection.
    \item Evaluating the performance of state-of-the-art retrieval models for this task and comparing them against the proposed method. 
\end{itemize}

Our analyses and experiments suggest specific future directions in this research area. 

\section{Related Work}
\label{sec:rel}

While the study of unified mobile search is a new research area, it has roots in previous research. 
Our work is related to the areas of mobile IR, federated, and aggregated search. Moreover, relevant research has been done in the area of proactive IR where a system aims to provide personalized information cards to users based on their context. Other relevant works can be found in the areas of query classification, neural networks, and crowdsourcing.
In the following, we summarize the related research in each of these areas.

\partitle{Mobile IR.} 
One of the main goals of mobile IR is to enable users to carry out all the classical IR operations using a mobile device~\cite{DBLP:series/sbcs/CrestaniMS17}.
One of the earliest studies on mobile IR was done by \citet{DBLP:conf/chi/KamvarB06} where they did a large-scale mobile search query analysis. They found mobile searches were less diverse in terms topic. 
In another study, \citet{DBLP:conf/mhci/ChurchSBC08} argued that the conventional Web-based approaches fail to satisfy users' information needs.
In fact, \citet{DBLP:conf/www/SongMWW13} found significant difference in search patterns done using iPhone, iPad, and desktop.
In a more recent study, \citet{DBLP:conf/sigir/Guy16} conducted an analysis on mobile spoken queries as opposed to typed-in queries. They found that spoken queries are longer and closer to natural language. These findings were in line with an older study by~\citet{DBLP:journals/jasis/CrestaniD06}. 

More recently, research has been done on various topics in mobile IR such as app and venue recommendation as well as app search~\cite{Park:2015,DBLP:conf/cikm/ParkFLZ16,DBLP:conf/sigir/AliannejadiC17}.
For instance, \citet{DBLP:conf/sigir/ShokouhiJORD14} studied query reformulation patterns in mobile query logs and found that users do not tend to switch between voice and text while reformulating their queries. 
\citet{Park:2015} represented apps using online reviews for improved app search on the market. 
\citet{DBLP:conf/www/WilliamsKCZAK16} leveraged mobile user gesture interactions, such as touch actions, to predict good search abandonment on mobile search.
\citet{DBLP:conf/cikm/ParkFLZ16} inferred users implicit intentions from social media for the task of app recommendation.
% \citet{DBLP:conf/sigir/OngJSS17} observed different user behaviors while doing mobile and desktop search as the amount of information scent was altered.
\citet{DBLP:conf/sigir/HarveyP17} found that fragmented attention of users while searching on-the-go, affects their search objective and performance perception. In contrast to the prior work, we explore how users behave while searching with different apps. To do this, we study the attributes of search queries assigned to different apps.

A few industrial systems exist aiming to provide users with unified mobile search. Apple Spotlight\footnote{\url{https://en.wikipedia.org/wiki/Spotlight_(software)}} is the most popular example of such systems that is available on iOS devices. Also, Sesame Shortcuts\footnote{\url{http://sesame.ninja/}} is an Android app that creates easy-to-access shortcuts to the installed apps. The shortcuts are also accessible via keyword-based queries. Despite the existence of these systems, research on cross-app search has not yet been done.

\partitle{Proactive IR.}
The aim of proactive IR systems is to anticipate users' information needs and proactively present information cards to them. \citet{DBLP:conf/sigir/ShokouhiG15} analyzed user interactions with information cards and found that the usage patterns of the cards depend on time, location, and user's reactive search history.
\citet{DBLP:conf/wsdm/BenetkaBN17} showed that information needs vary across activities as well as during the course of an activity. They proposed a method to leverage users' check-in activity for recommending information cards. Our work focuses on the queries that users issue in different apps. Queries can express complex information needs that are impossible to infer from context.

\partitle{Federated and aggregated search.} 
A unified mobile search system distributes a search query to a limited number of apps that it finds more relevant to a search query. There is a considerable overlap between the target apps selection task and federated/aggregated search.
In federated search, the query is distributed among uncooperative resources with homogeneous data; whereas in aggregated search, the content is blended from cooperative resources with heterogeneous data~\cite{DBLP:journals/ftir/Arguello17}. 
Given the uncooperative environment of most federated search systems, \citet{DBLP:journals/tois/CallanC01} proposed a query-based sampling approach to \textit{probe} various resource providers and modeled them based on the returned results. In most aggregated search systems, on the other hand, different resources are parts of a bigger search system and thus cooperative. Moreover, an aggregated search system can even access other metadata such as users' queries and current traffic~\cite{DBLP:journals/ftir/Arguello17}. \citet{DBLP:conf/wsdm/Diaz09} proposed modeling the query dynamics and collection to detect news queries for integrating the news \textit{vertical} into the result page. This work was later extended by~\citet{DBLP:conf/cikm/ArguelloCD09} to include images, videos, and travel information. 
In this work, we assume an uncooperative environment because the contents of apps are not accessible to the unified search system. Moreover, given the existence of various content types in different apps, we assume the documents to be heterogeneous.

\partitle{Query classification.}
Our work is also related to the research in query classification where different strategies are taken to assign a query to predefined categories.
\citet{DBLP:conf/sigir/KangK03} defined three types of queries arguing that search engines require different strategies to deal with the queries belonging to each of the classes. \citet{DBLP:conf/sigir/ShenSYC06} introduced an intermediate taxonomy used to classify queries to specified target categories. \citet{DBLP:conf/sigir/CaoHSJSCY09} leveraged conditional random fields to incorporate users' neighboring queries in a session as context. More recently, \citet{Zamani:2016} studied word embedding vectors for the query classification task and proposed a formal model for query embedding estimation. %In contrast, in this work, we aim at producing a ranked list of mobile apps that would process the query. Also, the queries are not limited to Web search and are on various types of information need.

% Our work is also related to the research in query classification where different strategies are taken to assign a query to predefined categories.
% \citet{DBLP:conf/sigir/KangK03} defined three types of queries arguing that search engines require different strategies to deal with the queries belonging to each of the classes. In 2005, ACM KDD organized a competition asking the contestants to classify the search queries by real internet users into 67 predefined categories~\cite{DBLP:journals/sigkdd/LiZD05}. \citet{DBLP:conf/sigir/ShenSYC06} later outperformed the top systems of the competition. They built a bridging classifier that would classify the query to an intermediate taxonomy used to classify the queries to the target categories. \citet{DBLP:conf/sigir/CaoHSJSCY09} leveraged conditional random fields to incorporate users' neighboring queries in a session as context. More recently, \citet{Zamani:2016} studied word embedding vectors for the query classification task and proposed a formal model for query embedding estimation. In contrast, in this work, we aim at producing a ranked list of mobile apps that would process the query. Also, the queries are not limited to Web search and are on various types of information need.

\partitle{Neural IR.}
The recent and successful development of deep neural networks for various tasks has also impacted IR applications. In particular, neural ranking models have recently shown significant improvements in a wide range of IR tasks, such as ad-hoc retrieval \cite{Guo:2016}, question answering \cite{Yu:2014}, and context-aware retrieval~\cite{Zamani:2017:WWW}. These approaches often rely on learning high-dimensional dense representations that carry semantic information. They can be particularly useful to match queries and documents where minimal term overlap exists. We also take advantage of such latent high-dimensional representations in our models for representing mobile apps.

% \partitle{Industrial systems.} 
% A few industrial systems exist aiming to provide users with unified mobile search. Apple Spotlight\footnote{\url{https://en.wikipedia.org/wiki/Spotlight_(software)}} is the most popular example of such systems that is available on iOS devices. Also, Sesame Shortcuts\footnote{\url{http://sesame.ninja/}} is an Android app that creates easy-to-access shortcuts to the installed apps. The shortcuts are also accessible via keyword-based queries. Despite the existence of these systems, research on cross-app search has not yet been done.

\partitle{Crowdsourcing.}
Although there has been a large body of research in IR related to crowdsourcing, we can only mention the most relevant works. 
\citet{DBLP:conf/sigir/AlonsoS14} explored building a query log that is coupled with query annotations describing the search tasks. \citet{DBLP:conf/ecir/ArguelloAD16} used crowdsourcing to collect spoken queries for predefined search tasks. In our work, we combine these two approaches (see Section~\ref{sec:col}).

\section{Crowdsourcing}
\label{sec:col}

\begin{table}
    \centering
    \caption{Distribution of crowdsourcing search task categories.}
    \label{tab:tasks}
    \begin{tabular}{ll}
    \toprule
        \textbf{Search Category} & \textbf{\% of tasks} \\
    \midrule
         General Information \& News & 13\% \\
         Video \& Music & 12\% \\
         Image & 9\% \\
         Social Networking & 9\% \\
         App & 9\% \\
         File \& Contact & 8\% \\
         Online Shopping & 13\% \\
         Local Services \& Navigation & 15\% \\
         Email \& Event & 12\% \\
    \bottomrule
    \end{tabular}
\end{table}

In this section, we describe how we collected \textit{\unimobile}, which is, to the best of our knowledge, the first dataset on cross-app mobile search queries. We started by creating a number of Human Intelligence Tasks (HITs) on Amazon Mechanical Turk\footnote{\url{http://www.mturk.com}}, asking workers to describe their latest mobile search experience in detail. The answers helped us to define fine-grained diverse naturalistic mobile search tasks. Then, we launched another task asking workers to assume they wanted to complete a given search task on their smartphones. They had to submit their search queries as well as the apps they would choose to complete each task.

\partitle{Task definition.} In the first crowdsourcing task, we described the category of search, giving them a handful of general examples. Furthermore, we also asked them to give us the context and background of their search, as well as the queries and the apps they used to do the search. Finally, we provided a complete example of a valid answer. We launched this job for most of search categories listed in Table~\ref{tab:tasks}. The HIT payment was \$0.10 and the workers were based in the U.S. with an overall acceptance rate of 75\% or higher.
The average work time was 246 seconds with 135 workers completing 169 HITs resulting in an average of 92 terms per HIT. The workers provided enough details about the context and background of their search that enabled us to generalize the task to the level that we would get a wide range of queries on the same task. For example, one worker submitted the following answer:
\begin{displayquote}
``I was searching for a new refrigerator to buy. The first thing I did was search for the best refrigerators of 2017 and then narrow down my search for exactly the type of refrigerator that I was looking for...''
\end{displayquote}
Then, we used this answer to define a more general search task:
\begin{displayquote}
``Consider one of the oldest appliances in your home. You have been thinking of changing it for a while. Now, it's time to order it online.''
\end{displayquote}

\begin{figure}[t]
    \centering
    {
        \setlength{\fboxsep}{0pt}
        \setlength{\fboxrule}{1pt}
        \fbox{\includegraphics[width=\columnwidth]{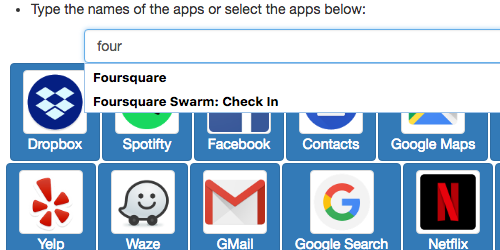}}%
    }
    \caption{HIT interface for choosing apps. The workers could enter an app's name or click on an app's icon.}
    \label{fig:mturk}
\end{figure}

\partitle{Query and app pairs.} The second crowdsourcing task consisted of 206 individual search task descriptions, mostly extracted from the answers we got in the first task. Table~\ref{tab:tasks} lists the distribution of the tasks. In the definition of tasks, our aim was to cover various aspects of mobile information seeking as mentioned in~\cite{DBLP:conf/iui/ChurchS09}. 
% It is worth noting that we also randomly selected and used 15 search tasks from the TREC Robust track~\cite{DBLP:journals/forum/Voorhees}. 
% Here the HITs started with an introduction of the task, followed by the steps workers had to follow. 
We asked the workers to read the search task description very carefully and assume that they wanted to perform it using their own mobile device. Then, we asked them to select one or more apps from a given list. Alternatively, they could type the name of the app they would choose for that search task. 
We provided an auto-complete feature for entering the apps' names in order to make it easier for the users to type the name of their favorite apps. Figure~\ref{fig:mturk} shows the interface we designed for this HIT.
Since we restricted the HIT to be done only by workers in the U.S., we chose the list of apps from the most popular Android apps in the U.S. market. 
Note that the apps were randomly shuffled and displayed to each worker to prevent any position bias.
These apps are listed as follows: Google Search, Gmail, Play Store, Facebook, Instagram, Google Maps, YouTube, Amazon, Twitter, Spotify, Waze, Pinterest, WhatsApp, File Manager, Netflix, Yelp, Contacts, Dropbox. 

As incentive, we paid \$0.05 for every HIT assignment. We also encouraged the workers to complete a survey for a \$0.05 bonus. Our aim was to understand the workers' background and familiarity with mobile devices. We asked the workers to perform the task using their mobile devices' browsers and tracked their keyboard keystrokes to prevent them from copying any text from the task description. The average work time for this task was 85 seconds with 91\% of the workers completing the survey. The key statistics of the survey were that 59\% of the workers used Android and 55\% used a mobile device as the primary device to connect to the Internet. Moreover, 83\% of the workers believed they use their mobile device more than two hours a day and 41\%, more than four hours a day. After launching several batches, we went through all the submitted answers for quality control and we observed that following crowdsourcing task design guidelines of~\cite{DBLP:conf/chi/KitturCS08} helped us achieve a very high assignment approval rate (99\%). 
We have made the collection publicly available for research purposes. The released data consists of the tasks that we defined through the first set of HITs as well as user queries in the second set of HITs, together with their corresponding ranked list of apps. The data can be used to study how users are engaged in searching with different apps. Also, the release of the defined tasks provides the opportunity to conduct a similar study in a lab setting on participants' mobile phones and compare the findings with this work.

\section{Data Analysis}
\label{sec:anl}
In this section, we present a thorough analysis of \unimobile, to understand how users issue queries in different apps, and which apps they choose to complete search tasks.
With the definition of 206 mobile search tasks, we were able to collect 5,812 search queries and their target apps. Overall, queries were assigned to 121 unique apps. Table~\ref{tab:stats} lists all the details of our dataset. In the following, we analyze different aspects of the data.

\begin{table}[t]
    \centering
    \caption{Statistics of \unimobile.}
    \label{tab:stats}
    \begin{tabular}{ll}
        \toprule
         \# queries & 5,812 \\
         \# unique queries & 5,567 \\
         \# users & 625 \\
         \# search tasks & 206 \\
         \# unique apps & 121 \\
         \# unique first apps & 70 \\
         \# unique second apps & 89 \\
         Mean unique apps per task & 7.51 $\pm$ 10.57 \\
         Mean query per user & 9.30 $\pm$ 20.30 \\
         Mean query per task & 28.21 $\pm$ 12.72 \\
         Mean query terms & 4.21 $\pm$ 2.45 \\
         Mean query characters & 24.83 $\pm$ 12.88 \\
         \bottomrule
    \end{tabular}
\end{table}

\begin{figure}
    \centering
    \subfloat[]{%
        \includegraphics[width=0.25\columnwidth,valign=t]{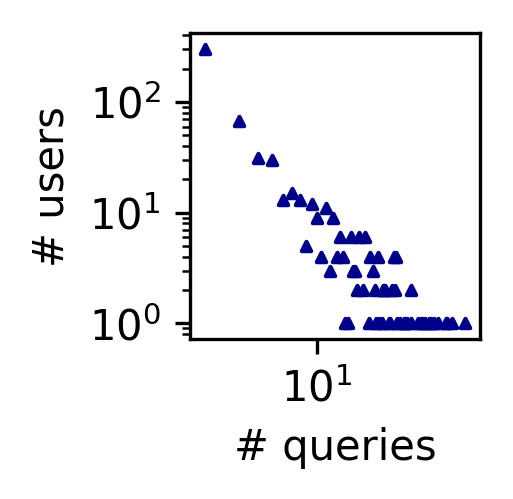}%
        \label{fig:user_count_count}
    }
    \subfloat[]{%
        \includegraphics[width=0.25\columnwidth,valign=t]{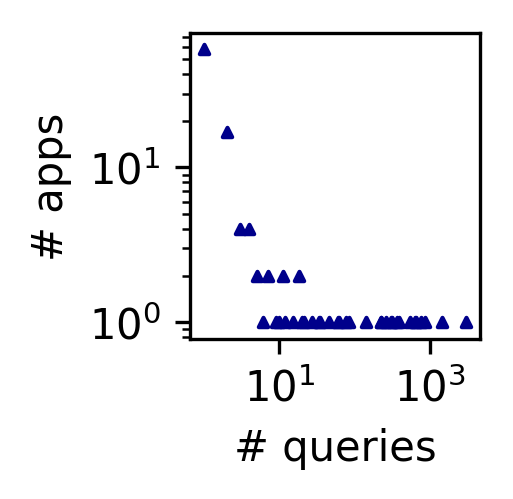}%
        \label{fig:app_count_count}
    }
    \subfloat[]{%
        \includegraphics[width=0.25\columnwidth,valign=t]{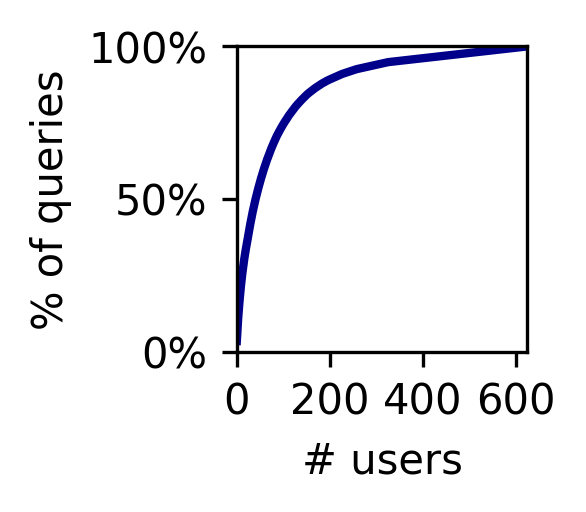}%
        \vphantom{\includegraphics[width=0.25\columnwidth,valign=t]{img/user_count_count_dist.png}}%
        \label{fig:user_cumm}
    } 
    \subfloat[]{%
        \includegraphics[width=0.25\columnwidth,valign=t]{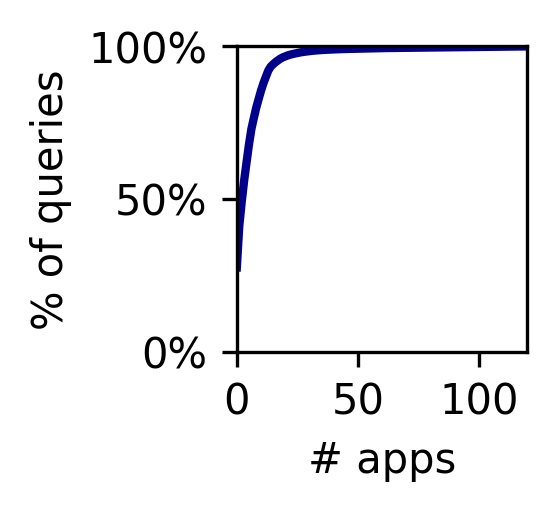}%
        \vphantom{\includegraphics[width=0.25\columnwidth,valign=t]{img/app_count_count_dist.png}}%
        \label{fig:app_cumm}
    }
    \caption{The distribution of number of queries with respect to apps and users.}
    % Despite the size of the figures, the presence of power law is evident.}
    \label{fig:app_user_dist}
\end{figure}

\begin{figure}
    \centering
    \includegraphics[width=0.8\columnwidth]{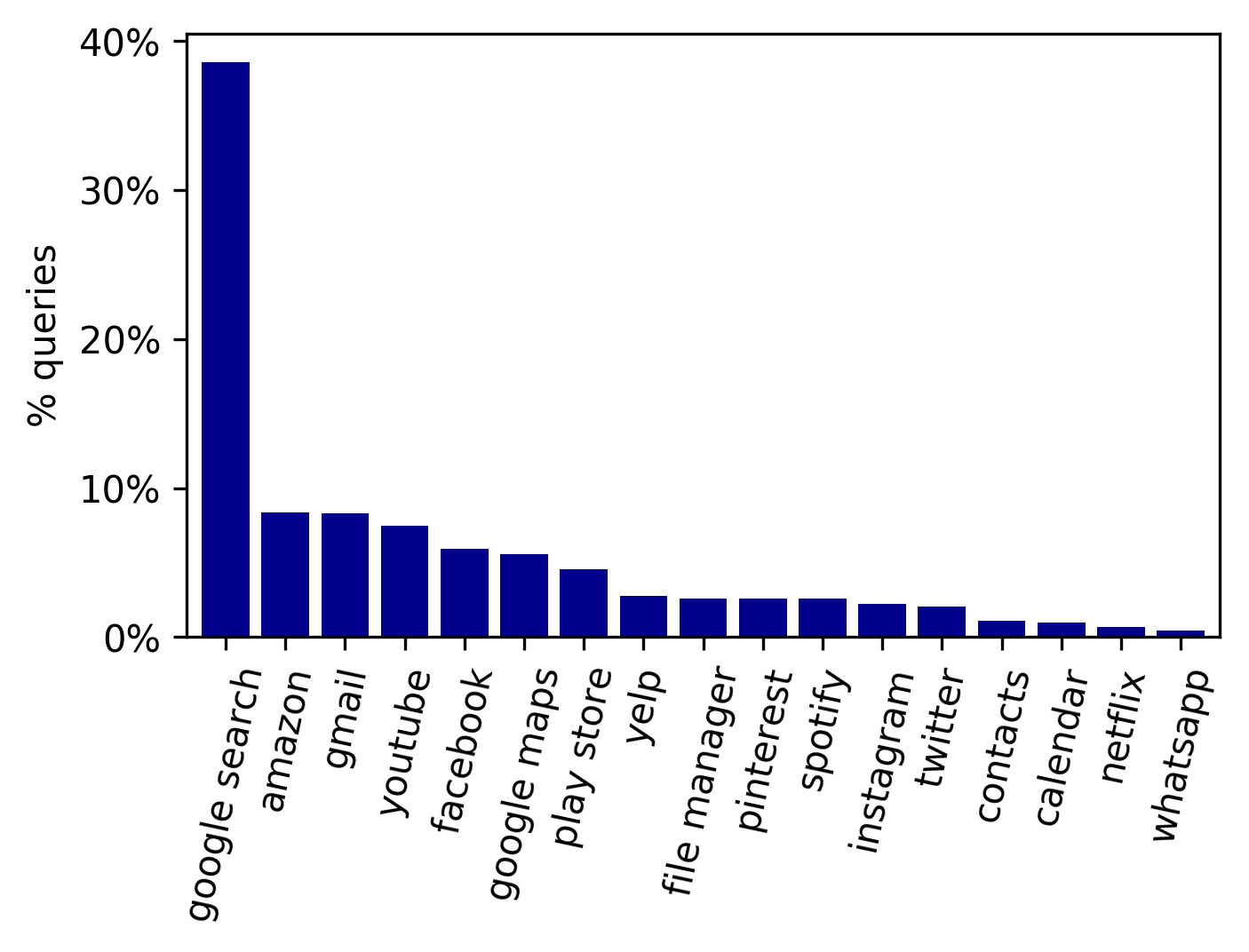}
    \caption{Number of queries per app for top 17 apps.}
    \label{fig:app_hist}
\end{figure}

\partitle{How apps are distributed.} Figure~\ref{fig:app_user_dist} shows the distribution of queries with respect to users and apps in \unimobile. 
As we can see in Figures~\ref{fig:user_count_count} and \ref{fig:user_cumm}, while there exist 173 users who submitted only one query, 110 users account for 80\% of the queries and 239 users account for 95\% of the queries. Also, we see in Figures~\ref{fig:app_count_count} and \ref{fig:app_cumm} that the distribution of apps follows a power-law. In particular, 9 apps account for more than 80\% and 17 apps account for more than 95\% of the queries. Figure~\ref{fig:app_hist} shows how queries are distributed with respect to the top 17 apps. As we can see, while Google Search\footnote{The term ``Google Search'' is also used to refer to the Google Chrome app.}, that is mainly targeted for Web search, constitute 39\% of total app selections, users opt to perform the majority (61\%) of their search tasks using other apps. 
Moreover, the variety of apps ranges from apps dealing with local phone data (e.g., Contacts and Calendar) to social media apps (e.g., Facebook and Twitter) indicating that they cover a wide range of search tasks. 

\begin{figure}
    \centering
    \subfloat[]{
        \includegraphics[width=0.5\columnwidth]{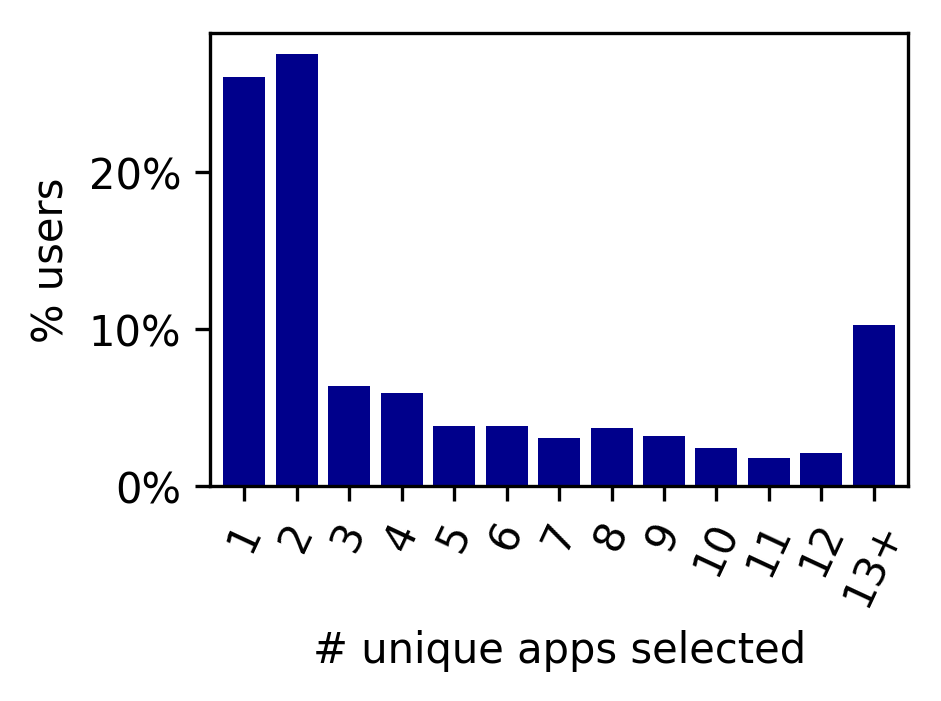}
        \label{fig:user_unique_app}
    }
    \subfloat[]{
        \includegraphics[width=0.5\columnwidth]{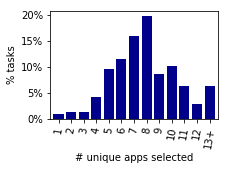}
        \label{fig:task_unique_app}
    }
    \caption{Distribution of unique apps per user and task.}
    \label{fig:unique_apps}
\end{figure}

\begin{figure*}
    \centering
    \subfloat[Google Search]{
        \includegraphics[width=0.13\textwidth]{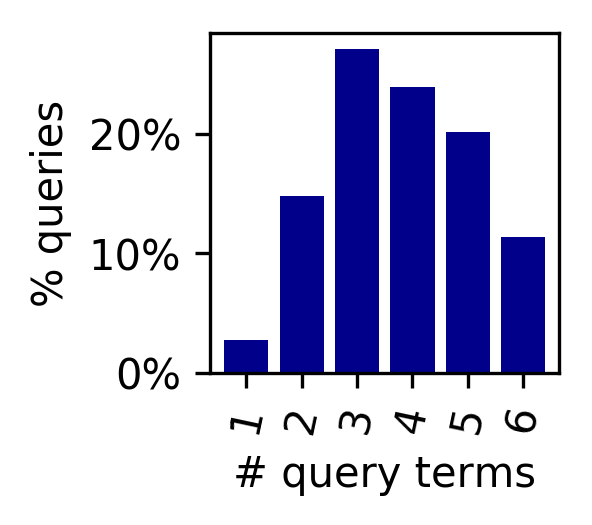}
        \label{fig:google_hist}
    }
    \subfloat[Gmail]{
        \includegraphics[width=0.13\textwidth]{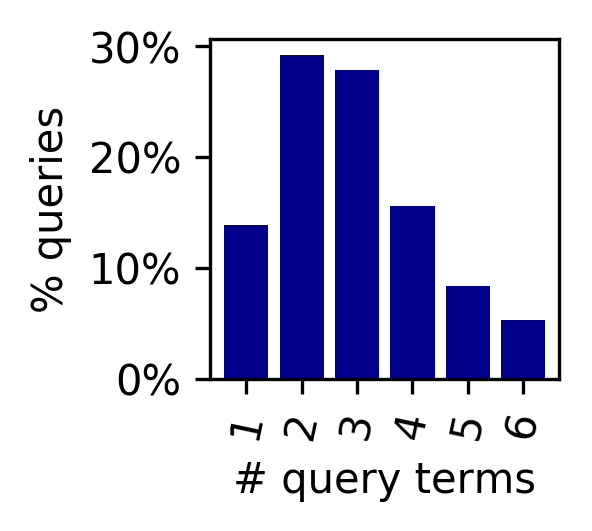}
        \label{fig:gmail_hist}
    }
    \subfloat[Yelp]{
        \includegraphics[width=0.13\textwidth]{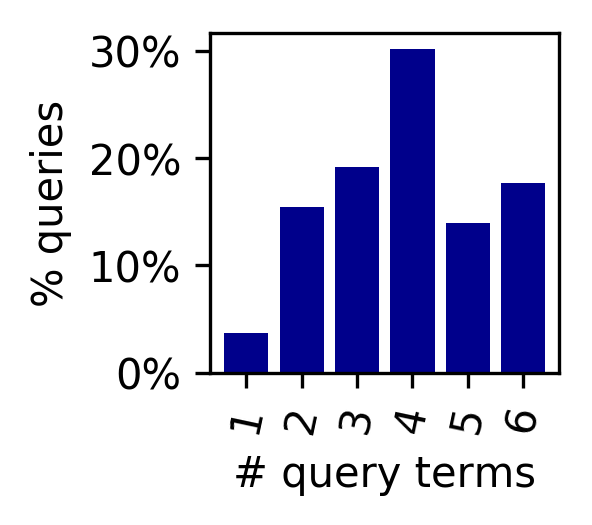}
        \label{fig:yelp_hist}
    }
    \subfloat[File Manager]{
        \includegraphics[width=0.13\textwidth]{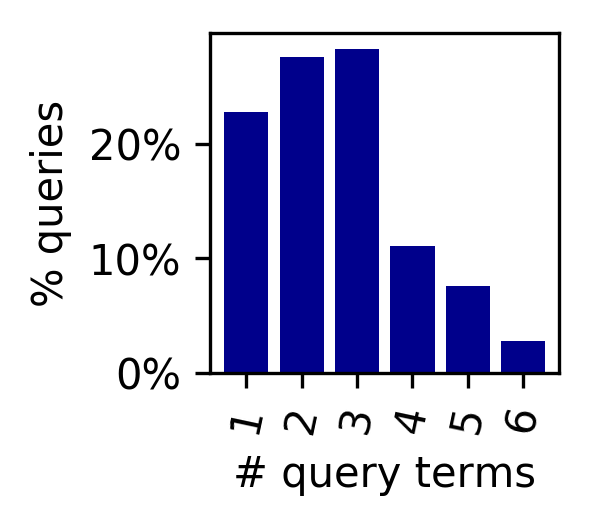}
        \label{fig:file_hist}
    }
    \subfloat[Contacts]{
        \includegraphics[width=0.13\textwidth]{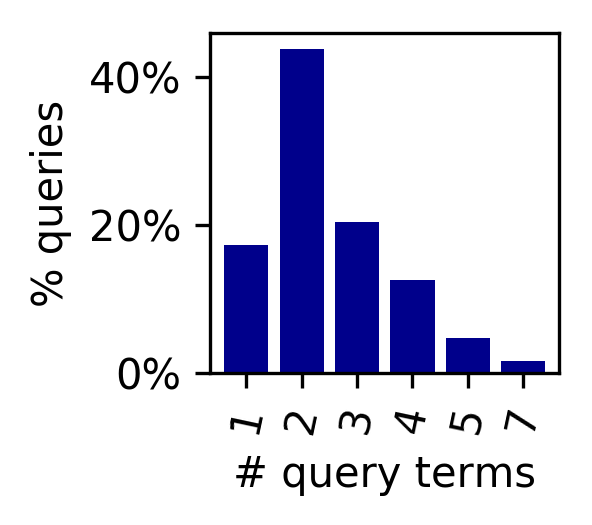}
        \label{fig:contacts_hist}
    }
    \subfloat[Calendar]{
        \includegraphics[width=0.13\textwidth]{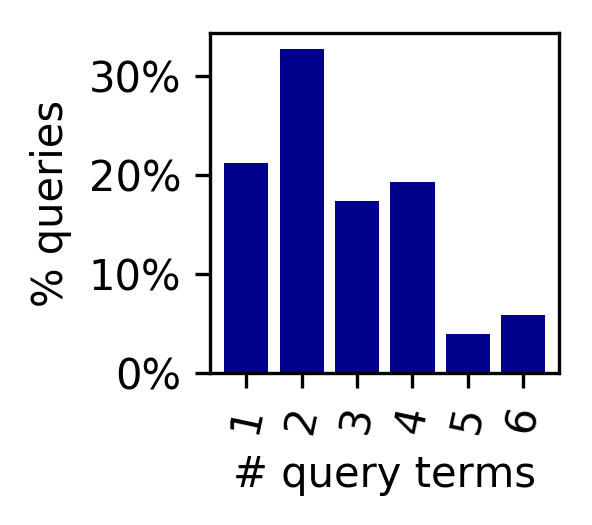}
        % \vphantom{\includegraphics[width=0.13\textwidth,valign=t]{img/whatsapp_query_len_hist.png}}
        \label{fig:calendar_hist}
    }
    \subfloat[WhatsApp]{
        \includegraphics[width=0.13\textwidth]{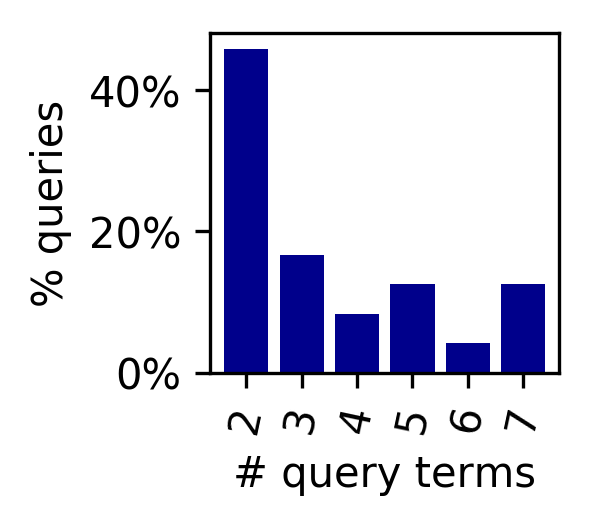}
        \label{fig:whatsapp_hist}
    }
    \caption{Histogram of number of query terms per app. Despite the small size, we can see the radically different distributions.}
    \label{fig:query_len_app}
\end{figure*}

\begin{figure}
    \centering
    \subfloat[]{
        \includegraphics[width=0.5\columnwidth,valign=t]{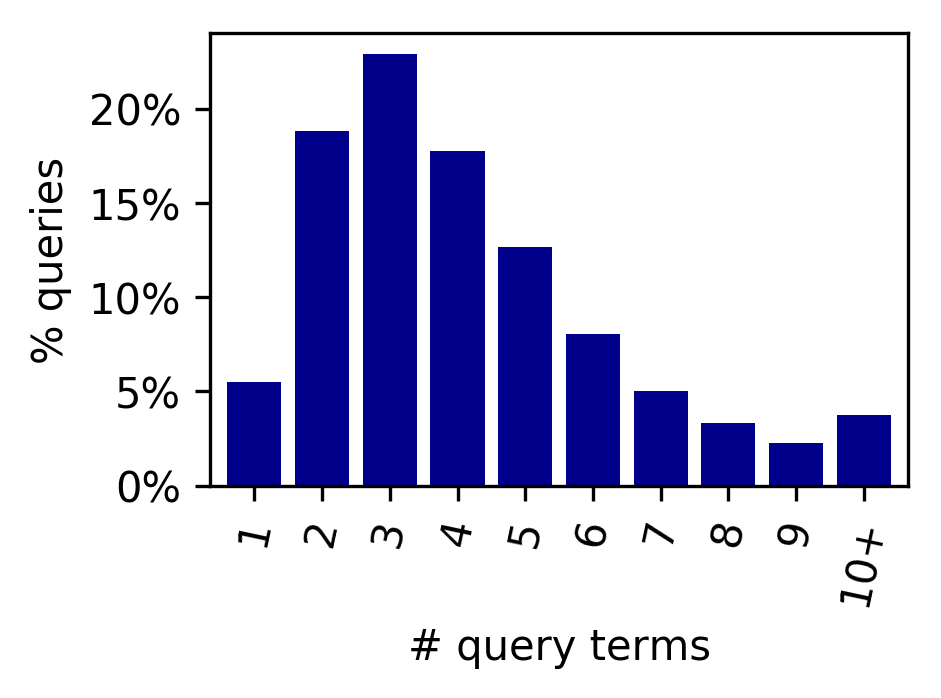}
        \label{fig:query_len_hist}
    }
    \subfloat[]{
        \includegraphics[width=0.5\columnwidth,valign=t]{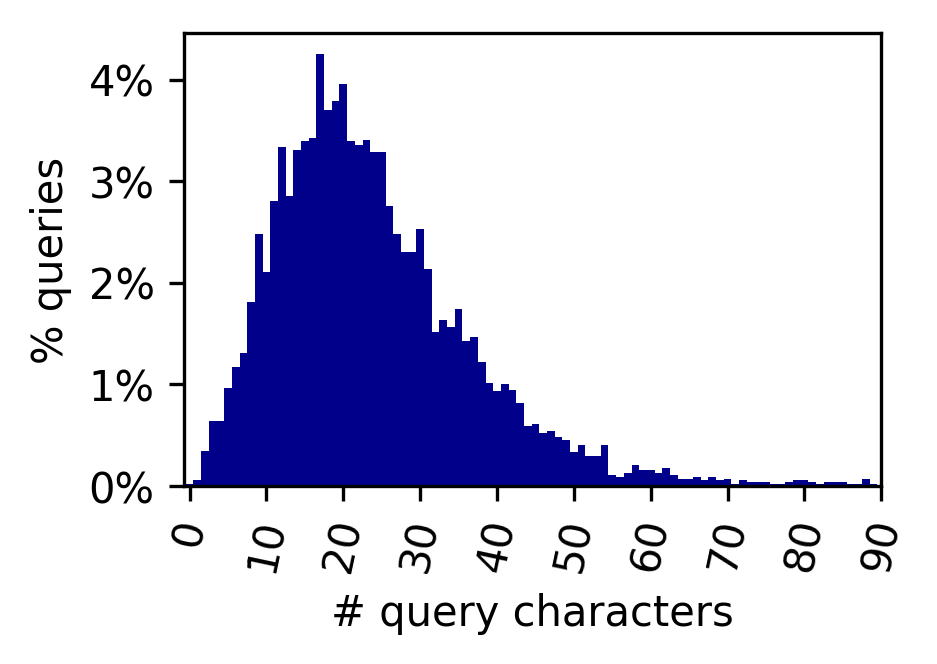}
        \vphantom{\includegraphics[width=0.5\columnwidth,valign=t]{img/query_term_hist.png}}
        \label{fig:query_char_hist}
    }
    \caption{Query length distribution with respect to number of terms and characters.}
    \label{fig:query_hist}
\end{figure}

% [distribution of queries for apps and related attributes]
\partitle{How apps are selected.} Here we are interested in finding out how users behave while choosing an app to perform a search task. Although users assign two apps while submitting 72\% of the queries, they choose only one app for 21\% of the queries and choose more than two apps for only 7\% of the queries. We also analyze how many different apps users select while doing the tasks. Figure \ref{fig:user_unique_app} shows the distribution of unique apps per users illustrating how many users selected a certain number of different apps. As we can see, a quarter of users preferred to search using two unique apps. On the other hand, Figure~\ref{fig:task_unique_app} plots the same distribution with respect to the tasks, that is how many unique apps were selected for each task. We see an entirely different distribution where the average number of unique apps per task is $7.51$, showing that the given search tasks can be addressed using multiple apps. As we compare the two distributions in Figures~\ref{fig:user_unique_app} and \ref{fig:task_unique_app}, we can conclude that while the majority of search tasks can be addressed using multiple apps, users usually limit their choice to a personal selection of apps. Therefore, a system can define a set of candidate apps which then can be narrowed down considering user's personal preference.

Furthermore, we analyze users' choice of Google Search, observing that it is selected as the first app in 39\% of the queries while 46\% as the second app. The users chose Google Search as the third app in 30\% of the queries with three selected apps. This indicates that, according to \unimobile, in most cases (61\%), users prefer to open a more specific app than Web search apps such as Google Search.
We also analyze users collective app selection behavior with respect to the tasks. For each task, we count how often each app is selected and sorted them. Our aim is to find out how often users decide to perform their search tasks using Google Search. According to our study, in 14\% of the tasks, no user selected Google Search, while in 35\% of the tasks Google Search was the most selected app. Moreover, in 68\% of the tasks it was among the top two most frequently selected apps, and in 78\% of the tasks it was among the top three. Considering the categorical distribution of apps in Table~\ref{tab:tasks} where only 13\% of the tasks were in the category of General Information \& News, we see that Google Search attracts many queries from the tasks that can be done using a more specific app. Given the integrity and aggregation of various search services such as image, video, location, and online shopping and easy access to them in one app, this observation is not surprising. Nevertheless, we see that for 86\% of the tasks, most users prefered other apps. 
This suggests that a unified mobile search system has a high potential of simplifying and improving users search experience.

% [query specifics]
\partitle{How queries differ among apps.} We analyze different attributes of queries with respect to their corresponding apps, to understand how different users formulate their information needs into queries using different apps. After tokenizing the queries, the average query terms per query is 4.21. We analyze the distribution of number of query terms per app, observing different distributions for every app, some of which are shown in Figure~\ref{fig:query_len_app}. This difference is more obvious if we compare Google Search with personal or local apps such as Contacts. In particular, Google Search has an average of 4.82 query terms while Contacts has an average of 2.67, which is considerably less than other apps. This can be explained if we consider the type of information users usually look for using Contacts app. The queries usually consist of one of the stored names on the phone, followed by terms such as ``email,'' ``address,'' ``info,'' and ``contact.'' Moreover, Figure~\ref{fig:query_hist} plots the distribution of query length with respect to terms and characters on the whole dataset.

\begin{figure}
    \centering
    \subfloat[Google Search]{
        \includegraphics[width=0.48\columnwidth,valign=t]{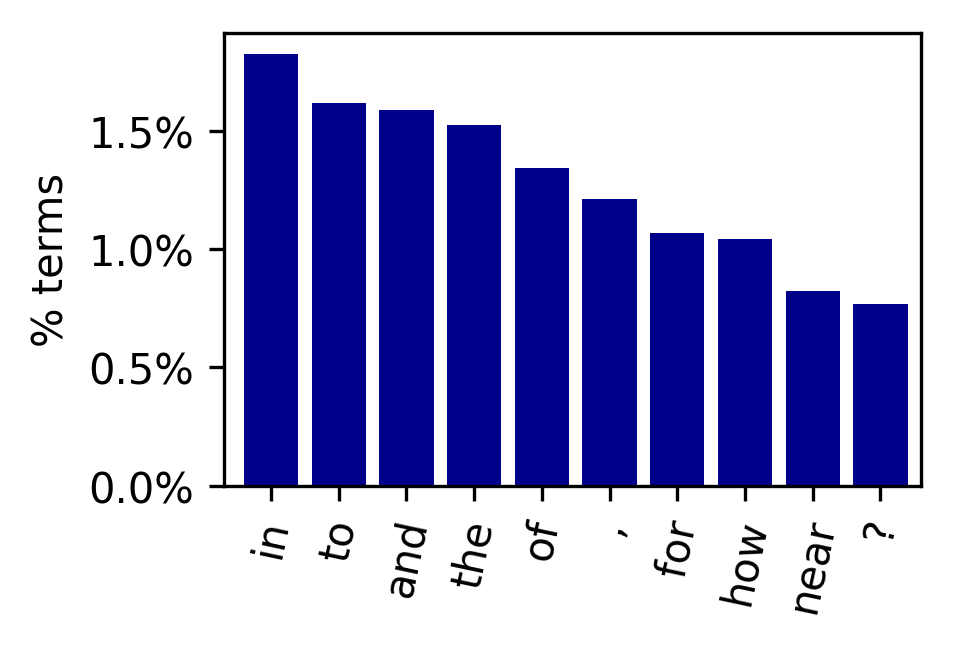}
        \vphantom{\includegraphics[width=0.48\columnwidth,valign=t]{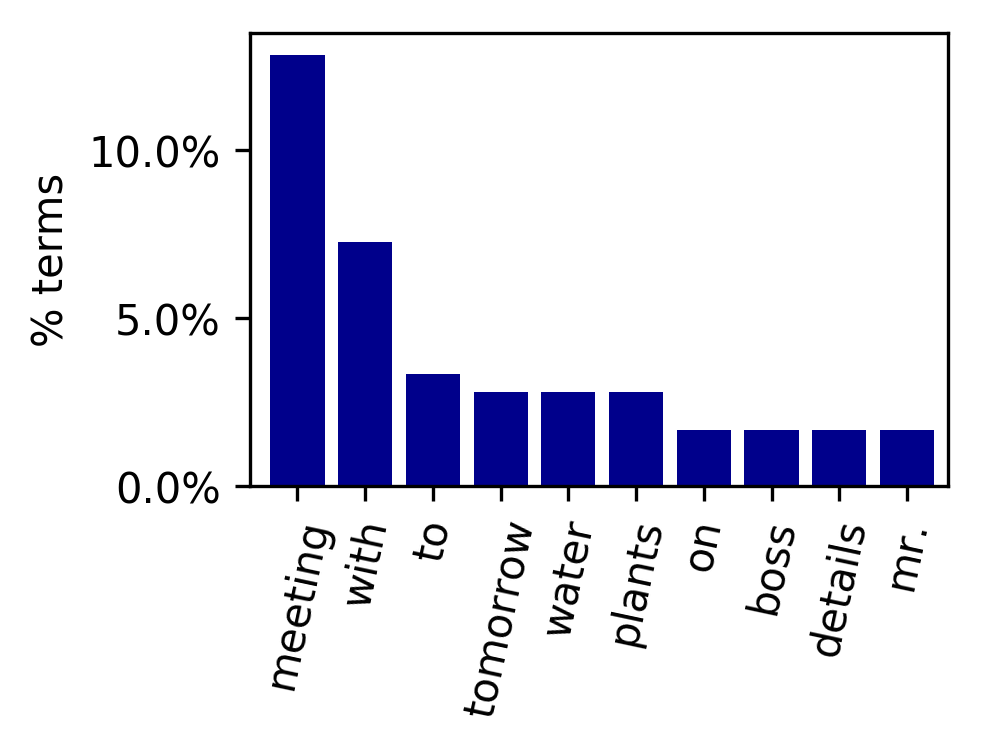}}
        \label{fig:google_terms}
    }
    \subfloat[Calendar]{
        \includegraphics[width=0.48\columnwidth,valign=t]{img/calendar_query_term_dist.png}
        \label{fig:calendar_terms}
    }
    \caption{Distribution of top query unigrams for two sample apps.}
    \label{fig:app_unigram}
\end{figure}

Figure~\ref{fig:query_len_app} demonstrates the distribution of number of query terms for 7 apps. In this figure, we only include the apps that exhibit a considerably different distribution from the average. As shown, Google Search query terms peak at 3 while personal apps such as Contacts, Calendar, and Gmail peak at 2. This indicates that the structure of queries vary depending on the target app. We can also see the difference in the most frequent unigrams for two example apps in Figure~\ref{fig:app_unigram} where we see that while stopwords are the most frequent unigrams used in queries submitted to Google Search, for a specific personal app such as Calendar, a domain-specific term such as ``meeting'' accounts for more than 15\% of the total distribution. This suggests that while considering domain-specific terms is crucial to predicting the target app, taking into account the query structure is also important. For instance, as we see in Figure~\ref{fig:google_terms}, the question mark is among the top query unigrams submitted to Google Search, suggesting that many of the queries are submitted in the form of a question. In contrast, as we mentioned earlier, the structure of contact queries are mostly in the form of ``<proper noun> + <information field>,'' as in ``sam email.''

\partitle{Query overlap.} Here we study query overlap or query similarity over the queries using a simple function used in previous studies done on large-scale query logs (e.g., \cite{DBLP:journals/tweb/ChurchSCB07}). We measure the query overlap at various degrees and use the similarity function $\text{sim}(q_1, q_2) = |q_1 \cap q_2|/|q_1 \cup q_2|$.
This function simply measures the overlap of query terms. We observed 70\% of queries overlapping with at least another query at the similarity threshold of $>0.25$. Higher thresholds lead to significantly lower similar queries; with thresholds $> 0.50$ and $> 0.75$ we observe that 24\% and 9\% of queries were similar, respectively. 
Similar to previous analyses, in Table~\ref{tab:query_overlap} we observe a different level of query overlap in queries associated with different apps.
The least query overlap is observed for Facebook queries. This could be due to the personal environment of Facebook. The highest query overlap is observed in Play Store queries. We observed the presence of some domain-specific terms such as ``app'' in many queries which results in higher query similarity. The observed difference in query overlap for every app suggests that various factors influence the way users formulate their queries. For example, apps that provide more focused information, receive more similar queries. On the other hand, more personal apps receive a diverse set of queries as they reflect personal information needs which can be totally different from one user to the other.

\partitle{Summary.} 
Our analyses first showed that users' queries are mainly targeted to a few apps; however, these apps are very different in terms of their content. Moreover, we showed that users often choose two different apps for a single query, suggesting that many users submit the same query in multiple apps. Also, we showed that different users select an average of more than 7 apps for each task, with Google Search being the top selected app in only 35\% of the cases. This again indicates the necessity of a unified search system on mobile devices. Finally, we analyzed the queries issued in different apps and found notable differences. For instance, we showed that query lengths, unigram distribution, and query overlap differ among apps. This suggests that the query structure needs to be taken into account while representing the apps.

\section{Neural Target Apps Selection}
\label{sec:method}
Assume that a user aims at submitting a query $q$ to a set of mobile apps $\{a_1, a_2, \cdots, a_n\}$, called the target apps. Note that the size of this set could be equal to $1$. The task of \emph{target apps selection} is defined as ranking the mobile apps in response to the query $q$, such that the target apps appear in higher ranks. In this section, we propose our methodology to tackle the target apps selection task. To this end, we propose two \emph{general} frameworks based on neural networks. Our first framework, called \ntar1, is given a query and a candidate app and produces a retrieval score. We study both pointwise and pairwise training settings for this framework. Our second framework, called \ntar2, is given a query as the input and produces a probability distribution indicating the probability of each app being targeted, for all apps. 

One of the main challenges in this task is that it is not obvious how to represent each app. For example, although the apps' descriptions would be used for app representation in the app selection task \cite{Park:2015}, it cannot be used in the target apps selection. Because the queries that can be searched in a specific app do not match with the content of the app's description. To address this issue, our frameworks learn a high-dimensional representation for each app, as part of the network. The following subsections describe these two frameworks in more detail.

\begin{table}[t]
    \caption{The percentage of similar queries at different similarity thresholds considering only the queries associated with every app.}
    \label{tab:query_overlap}
    \centering
    \begin{tabular}{lccc}
        \toprule
         \multirow{2}{*}{\textbf{App}} & \multicolumn{3}{c}{\textbf{\% of similar queries}}  \\
         \cline{2-4}
         & $> 0.25$ & $> 0.50$ & $> 0.75$  \\
         \midrule
          All apps & 70\% & 24\% & 9\% \\
          Google Search & 63\% & 19\% & 6\% \\
          Amazon & 38\% & 8\% & 3\% \\
          Gmail & 57\% &  14\% & 7\% \\
          YouTube & 49\% & 20\% & 7\% \\
          Google Maps & 46\% & 3\% & 1\% \\
          Facebook & 30\% & 9\% & 1\%\\
          Play Store & 61\% & 26\% & 14\% \\
        \bottomrule         
    \end{tabular}
\end{table}

\subsection{\ntar1: App Scoring Model}
\ntar1~outputs a retrieval score for a given query $q$ and a candidate app $a$. Formally, \ntar1 can be defined as follows:
\begin{equation}
    \text{score} = \psi(\phi_Q(q), \phi_A(a))~,\nonumber
\end{equation}
where $\psi(\cdot, \cdot) \in \mathbb{R}$ is a scoring function for the given query representation $\phi_Q(q) \in \mathbb{R}^m$ and app representation $\phi_A(a) \in \mathbb{R}^{n}$. Various neural architectures can be employed to model each of the three components in the \ntar1~framework.

We implement the component $\phi_Q(q)$ with two major functions: an embedding function $\mathcal{E} : V \rightarrow \mathbb{R}^{d}$ that maps each vocabulary term to a $d$-dimensional embedding space, and a global term weighting function $\mathcal{W}: V \rightarrow \mathbb{R}$ that maps each vocabulary term to a real-valued number showing its global importance. The query representation function $\phi_Q$ represents a query $q = \{w_1, w_2, \cdots, w_{|q|}\}$ as follows:
\begin{equation}
\phi_Q(q) = \sum_{i=1}^{|q|} \widehat{\mathcal{W}}(w_i)\cdot \mathcal{E}(w_i)~,
\label{eq:queryrep}
\end{equation}
which is the weighted element-wise summation over the terms' embedding vectors (hence, $m=d$). $\widehat{\mathcal{W}}$ is the normalized global weights computed using a softmax function as follows:
\begin{equation}
\widehat{\mathcal{W}}(w_i) = \frac{\exp(\mathcal{W}(w_i))}{\sum_{j=1}^{|q|}{ \exp(\mathcal{W}(w_j))}}~.\nonumber
\end{equation}

This is a simple yet effective approach for query representation based on the bag of words assumption, which has been proven to be effective for the ad-hoc retrieval task \cite{Dehghani:2017}. Note that the matrices $\mathcal{E}$ and $\mathcal{W}$ are the network parameters in our model and are learned to provide task-specific representations.

The app representation component $\phi_A$ is simply implemented as a look-up table. In other words, our neural model consists of an app representation matrix $\mathcal{A} \in \mathbb{R}^{N \times n}$ where $N$ denotes the total number of apps and the $i$\textsuperscript{th} row of this matrix is a $n$-dimensional representation for the $i$\textsuperscript{th} app. Therefore, $\phi_A(a)$ returns a row of the matrix $\mathcal{A}$ that corresponds to the app $a$.

To model the function $\psi$, following \citet{Zamani:2018}, we feed the Hadamard product (which enforces $m=n$) of the learned query and app representations into a fully-connected feed-forward network with two hidden layers. This network produces a single output as the score assigned to the given query-app pair. We use rectified linear unit (ReLU) as the activation function in the hidden layers of the network. To prevent overfitting, the dropout technique \cite{Srivastava:2014} is employed.

We study both pointwise and pairwise learning settings for our \ntar1~model.

\partitle{Pointwise learning.} In a pointwise setting, we use mean squared error (MSE) as the loss function. MSE for a mini-batch $b$ is defined as follows:
\begin{equation}
    \mathcal{L}_{MSE}(b) = \frac{1}{|b|}\sum_{i=1}^{|b|}{(y_i - \psi(\phi_Q(q_i), \phi_A(a_i)))^2}~,\nonumber
\end{equation}
where $q_i$, $a_i$, and $y_i$ denote the query, the candidate app, and the label in the $i$\textsuperscript{th} training instance of the mini-batch. For this training setting, we use a linear activation for the output layer.

\partitle{Pairwise learning.} \ntar1~can be also trained using a pairwise setting. Therefore, each training instance consists of a query, a target app, and a non-target app. To this end, we employ hinge loss (max-margin loss function) that has been widely used in the learning to rank literature for pairwise models \cite{Li:2011}. Hinge loss for a mini-batch $b$ is defined as follows:
\begin{align}
    \mathcal{L}_{Hinge}(b) = \frac{1}{|b|}\sum_{i=1}^{|b|}&\max \left\{0, \epsilon-\right.\left.\text{sign}(y_{i1} - y_{i2}) \right. \nonumber\\
    &  \left. \left(\psi(\phi_Q(q_i), \phi_A(a_{i1})) - \psi(\phi_Q(q_i), \phi_A(a_{i2}))\right)\right\}~,\nonumber
\end{align}
where $\epsilon$ is a hyper-parameter determining the margin of hinge loss, a linear loss function that penalizes examples violating the margin constraint. To bound the output of the model to the $[-1, 1]$ interval, we use $\tanh$ as the activation function for the output layer, in the pairwise training setting. The parameter $\epsilon$ is also set to $1$, which works well when the predicted scores are in the $[-1, 1]$ interval.

\subsection{\ntar2: Query Classification Model}
Unlike \ntar1~that predicts a score for a given query-app pair, our second framework computes the probability of each app being targeted by a given query. In more detail, \ntar2~is modeled as $\gamma(\phi_Q(q)) \in \mathbb{R}^N$, whose $i$\textsuperscript{th} element denotes the probability of the $i$\textsuperscript{th} app being targeted, given the query representation $\phi_Q(q)$. $N$ is the total number of apps.

To implement \ntar2, we represent each query via a weighted element-wise average as explained in Equation~\eqref{eq:queryrep}. $\gamma$ is modeled using a fully-connected feed-forward network with the output dimension of $N$. ReLU is employed as the activation function in the hidden layers, and a softmax function is applied on the output layer to compute the probability of each app being targeted by the query.

To train \ntar2, we use a cross-entropy loss function which for a mini-batch $b$ is defined as:
\begin{equation}
    \mathcal{L}_{ce}(b) = \frac{1}{|b|}\sum_{i=1}^{|b|}{\sum_{j=1}^{N}(p(a_j | q_i) \log \gamma(\phi_Q(q_i)))}~.\nonumber
\end{equation}

Similar to \ntar1, we use dropout to regularize the model.

\section{Experiments}
\label{sec:exp}

\begin{table*}[]
    \centering
    \caption{Performance comparison with baselines on \unimobile-Q and \unimobile-T. The superscript * denotes significant differences compared to all the baselines.}
    \label{tab:results}
    \resizebox{17.9cm}{!}{%
    \begin{tabular}{lccccccccccc}
    \toprule
     \multirow{2}{*}{\textbf{Method}} & \multicolumn{5}{c}{\textbf{\unimobile-Q Dataset}} && \multicolumn{5}{c}{\textbf{\unimobile-T Dataset}} \\
     \cmidrule{2-6} \cmidrule{8-12} 
     & MRR & P@1 & nDCG@1 & nDCG@3 & nDCG@5 && MRR & P@1 & nDCG@1 & nDCG@3 & nDCG@5\\
    \midrule
    \textbf{StaticRanker} & 0.6485 & 0.5293 & 0.4031 & 0.4501 & 0.5144 && 0.6718 & 0.5507 & 0.4247 & 0.4853 & 0.5446 \\
    \textbf{QueryLM} & 0.5867 & 0.3803 & 0.3068 & 0.4676 & 0.5508 && 0.5178 & 0.3272 & 0.2619 & 0.3716 & 0.4503 \\ %
    \textbf{BM25} & 0.7523 & 0.6233 & 0.4915 & 0.6298 & 0.6859 && 0.6780 & 0.5244 & 0.4101 & 0.5392 & 0.5992 \\ %
    \textbf{BM25-QE} & 0.6948 & 0.5177 & 0.4116 & 0.5909 & 0.6498 && 0.6256 & 0.4276 & 0.3312 & 0.5015 & 0.5704 \\ %
    \textbf{k-NN} & 0.7373 & 0.6031 & 0.4794 & 0.6091 & 0.6633 && 0.6879 & 0.5414 & 0.4287 & 0.5413 & 0.6003 \\
    \textbf{k-NN-AWE} & 0.7420 & 0.6081 & 0.4842 & 0.6156 & 0.6682 && 0.6984 & 0.5551 & 0.4407 & 0.5560 & 0.6117 \\
    \textbf{LambdaMART} & 0.7313 & 0.6127 & 0.4864 & 0.6110 & 0.6426 && 0.6749 & 0.5469 & 0.4323 & 0.5419 & 0.5704 \\
    \midrule
    \textbf{\ntar1-pointwise} & 0.7591* & 0.6214 & 0.4897 & 0.6328 & 0.6934* & & 0.7047* & 0.5582* & 0.4493* & 0.5506* & 0.6258* \\
    \textbf{\ntar1-pairwise} & \textbf{0.7661}* & \textbf{0.6285}* & \textbf{0.5012}* & \textbf{0.6364}* & \textbf{0.7018}* & & \textbf{0.7192}* & 0.5661* & \textbf{0.4709}* & \textbf{0.5941}* & \textbf{0.6471} \\
    \textbf{\ntar2} & 0.7638* & 0.6271* & 0.4996* & 0.6351* & 0.6976* & & 0.7144* & \textbf{0.5723}* & 0.4608* & 0.5689* & 0.6334*\\
    % \midrule
    \bottomrule
    \end{tabular}
    }
\end{table*}

In this section, we evaluate the performance of the proposed models in comparison with a set of state-of-the-art IR models. We also study the performance of the models with respect to tasks and users.

\subsection{Experimental Setup}

\partitle{Dataset.} 
We evaluated the performance of our proposed models on the \unimobile~dataset. We followed two different strategies to split the data:
\begin{enumerate*}
    \item In \textit{\unimobile-Q}, we randomly selected 70\% of the \textit{queries} for training, 10\% for validation, and 20\% for test set
    \item In \textit{\unimobile-T}, we randomly split the tasks (rather than queries). To do so, we randomly selected 70\% of the \textit{tasks} for training, 10\% for validation, and 20\% for test set. 
\end{enumerate*} 
To minimize random bias, for each splitting strategy we repeated the process five times. The hyper-parameters of the models were tuned based on the results on the validation sets. Therefore, we repeated all the experiments five times and reported the average performance.

\partitle{Evaluation metrics.} 
Effectiveness was measured by five standard evaluation metrics: mean reciprocal rank (MRR), precision of the top 1 retrieved app (P@1), normal discounted cumulative gain for the top 1, 3, and 5 retrieved apps (nDCG@1, nDCG@3, nDCG@5). We determined the statistically significant differences using the two-tailed paired t-test with Bonferroni correction at a $95\%$ confidence interval ($p < 0.05$). 
In the ranked list of apps associated to every query, we assigned the score of 2 to the \textit{first} relevant app and 1 to the rest of relevant apps, to differentiate between a model that is able to rank the first relevant app higher and a model that is not.

The choice of evaluation metrics was motivated by considering three different aspects of the task, inspired by data analysis. We chose MRR considering scenarios where a user is looking for relevant information only in one app, and so they would stop scanning the search results as soon as they find the first relevant document. We reported P@1 and nDCG@1 to measure the performance for scenarios that a user only checks the first result.
Given that many search tasks need to be addressed using more than one app, it is crucial to evaluate a system with respect to more than one relevant app in the top-$k$ results.
nDCG@3 allowed us to evaluate our approach when a user scans the top 3 results. Since we found that most of the queries were assigned to one or two apps (see Section~\ref{sec:anl}), nDCG@3 measures how well a system is able to place the two relevant apps among the top 3 results. We also used nDCG@5 to evaluate top 5 results on a single screen, given the size of a typical smartphone. 

\partitle{Compared methods.} 
We compared the performance of our model with the following methods:
\begin{itemize}[leftmargin=*]
    % \item \textit{Random:} The scores of apps for a given query are the average of 100 random scores. As we saw in Figure~\ref{fig:app_cumm}, as small number of apps account for most of the dataset. Hence, we boosted the score of top-$k$ apps to be ranked higher.
    \item \textit{StaticRanker:} For every query we ranked the apps in the order of their popularity in the training set as a static (query independent) model.
    % than other apps.
    % and determined the value of $k$ using the validation set.
    \item \textit{QueryLM, BM25, BM25-QE:} For every app we aggregated all the relevant queries from the training set to build a document representing the app. Then we used Terrier~\cite{DBLP:conf/ecir/OunisAPHMJ05} to index the documents. QueryLM uses the language model retrieval model~\cite{DBLP:conf/sigir/PonteC98}. For BM25-QE, we adopted Bo1~\cite{DBLP:phd/ethos/Amati03} model for query expansion. We used the Terrier implementation of these methods.
    \item \textit{k-NN, k-NN-AWE:} To find the nearest neighbors in k nearest neighbors (k-NN), we considered the cosine similarity between TF-IDF vectors of queries. Then, we took the labels (apps) of the nearest queries and produced the app ranking. As for k-NN-AWE, we computed the cosine similarity between the average word embedding (AWE) of the queries obtained from GloVe~\cite{pennington2014glove} with 300 dimensions.
    \item \textit{LambdaMART:} For every query-app pair, we used the scores obtained by BM25, k-NN, and k-NN-AWE as features to train LambdaMART~\cite{DBLP:journals/ir/WuBSG10} implemented in RankLib\footnote{\url{https://sourceforge.net/p/lemur/wiki/RankLib/}}. For every query, we considered all irrelevant apps as negative samples.
\end{itemize}

\subsection{Results and Discussion}
In the following, we evaluate the performance of \ntar1 and \ntar2 trained on both data splits. We further analyze how other baseline models perform comparing their performance on both splits together with other methods.

\partitle{Performance comparison.} Table~\ref{tab:results} lists the performance of our proposed methods as well as the compared methods. As we can see, the performance of all methods drops when we use \unimobile-T data splits, except for StaticRanker. StaticRanker gives us an idea of how much the test set is biased towards more popular apps. For example, we see that StaticRanker performs better on \unimobile-T suggesting that it consists of more popular apps.
As we compare the relative performance drop between the two data splits, we see that among other baselines, k-NN-AWE is more robust with the minimum relative drop ($-$8.4\% on average). QueryLM, on the other hand, is the least robust model with the maximum relative drop ($-$16\% on average). This indicates that k-NN-AWE is able to capture similar queries for unseen tasks using a pre-trained word embedding, whereas QueryLM relies heavily on the indexed queries.

Among the baselines tested on \unimobile-Q, we see that BM25 performs best in terms of all evaluation metrics. Given that \unimobile-Q contains queries belonging to the same tasks both in training and test sets, this shows that when more similar queries exist in the index, BM25 is able to rank the apps more effectively.
However, on \unimobile-T, k-NN-AWE performs best in terms of all metrics. Given that \unimobile-T \textit{does not} contain queries belonging to the same task in training and test sets, this suggests that leveraging a pre-trained word embedding helps k-NN capture query similarities more effectively when the queries are less similar, leading to a better generalization. This can also be seen when comparing the performance of k-NN and k-NN-AWE, given that k-NN-AWE consistently outperforms k-NN. Regarding LambdaMART, we see that even though it benefits from multiple features, it does not perform as well as k-NN-AWE and BM25 on \unimobile-Q. On the contrary, we see that it performs better on \unimobile-T showing that the AWE-based feature improves its generalization.

As we can see, \ntar1-pairwise and \ntar2 outperform all the methods, on both data splits, in terms of all evaluation metrics. All the improvements are statistically significant suggesting that using queries to learn the app representation helps our approach learn the similarities more effectively. Considering the relative difference on the two data splits, we observe that our proposed approaches also show a drop. Compared to other methods (except for StaticRanker), we observe that \ntar1-pairwise and \ntar2 consistently have a lower relative drop across \unimobile-Q and \unimobile-T, indicating that the trained app embedding is an effective way to represent mobile apps based on the queries that are assigned to them. Among our proposed methods, \ntar1-pairwise has the least relative drop ($-$7.4\% on average), suggesting that a pairwise setting leads to a higher generalization.
%\hamed{needs further analysis}

\begin{figure}
    \centering
    \includegraphics[width=\columnwidth]{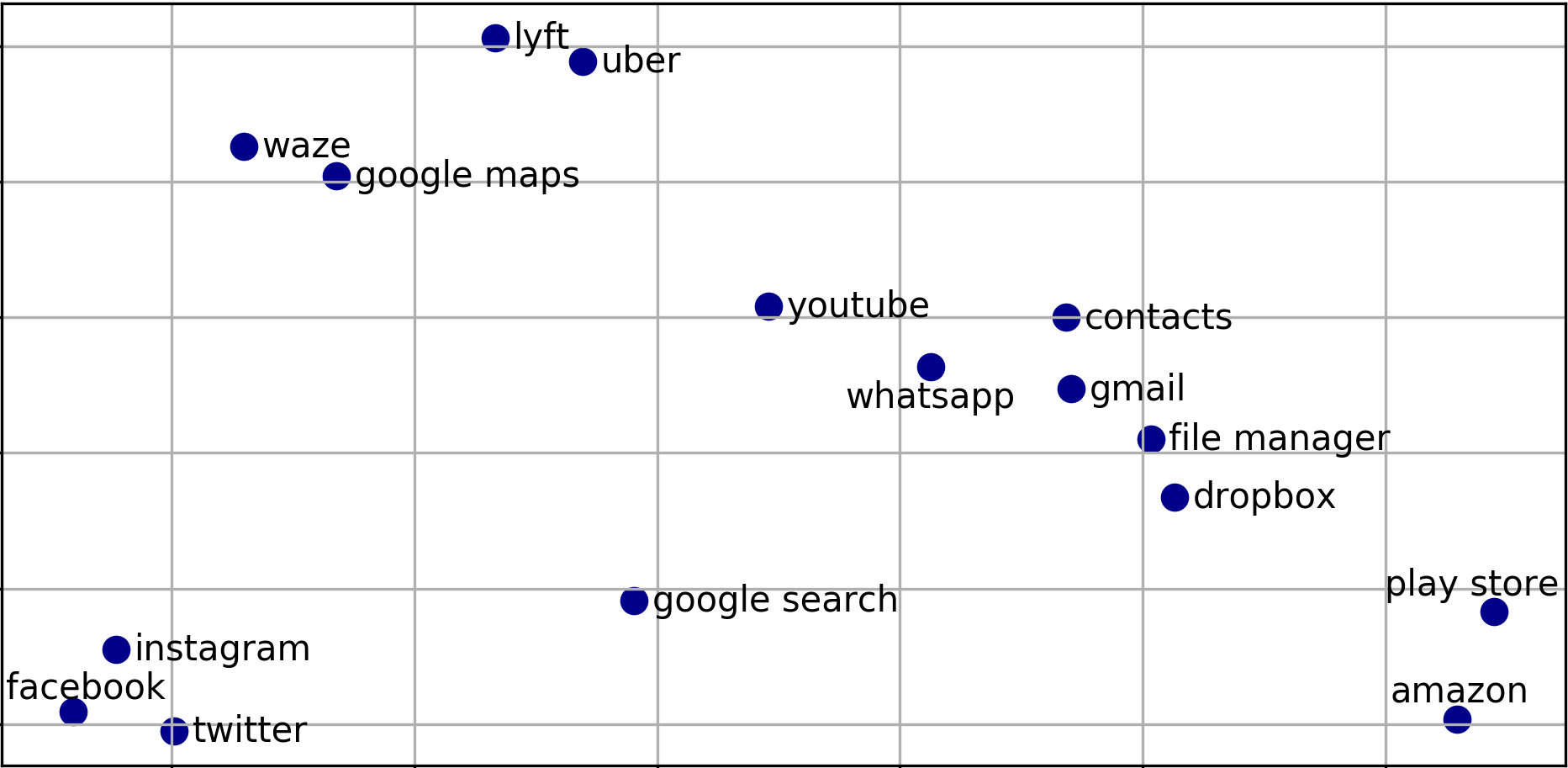}
    \caption{Proximity of different app representations learned by \ntar1 (pairwise). This plot is produced by reducing the dimensionality (using the t-SNE algorithm) of the app representations to two for visualization.}
    \label{fig:2d-rep}
\end{figure}

\partitle{Representation analysis.}
We reduce the dimensionality of the learned app representations by projecting them to a two-dimensional space using t-Distributed Stochastic Neighbor Embedding (t-SNE)~\cite{maaten2008visualizing}.
Figure~\ref{fig:2d-rep} shows the proximity of the representation of different apps\footnote{Given space limitations, we could not include all the apps in this figure.} being grouped in some clusters. For instance, all social media apps are placed close to each other. Also, we see that location search and navigation apps are in another cluster. Interestingly, Gmail is close to File Manager, Contacts, and WhatsApp. People usually search for attachments or their contacts using Gmail, explaining their proximity. Google Search, on the other hand, belongs to no cluster. This could be due to the variety of queries people submit to Google Search, placing it somewhere in the center of all other apps. 
However, we cannot explain why YouTube is close to WhatsApp, or why Play Store is close to Amazon.
Hence, Figure~\ref{fig:2d-rep} shows that learning high-dimensional app representation using the queries submitted to them is effective, though perhaps not perfect.

\begin{figure}
    \centering
    \subfloat[UniMobile-Q Dataset]{
        \includegraphics[width=0.99\columnwidth,valign=t]{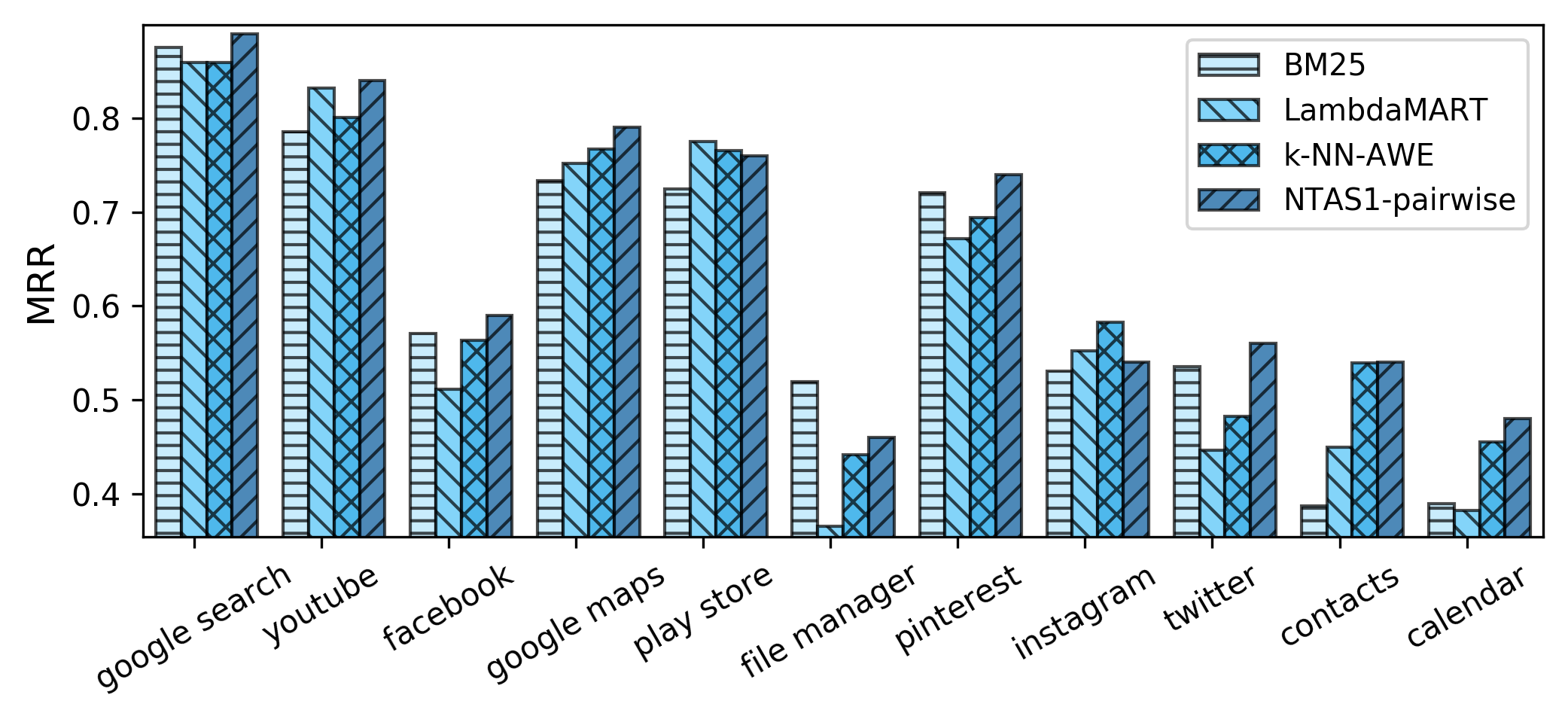}
    }
    
    \subfloat[UniMobile-T Dataset]{
        \includegraphics[width=0.99\columnwidth,valign=t]{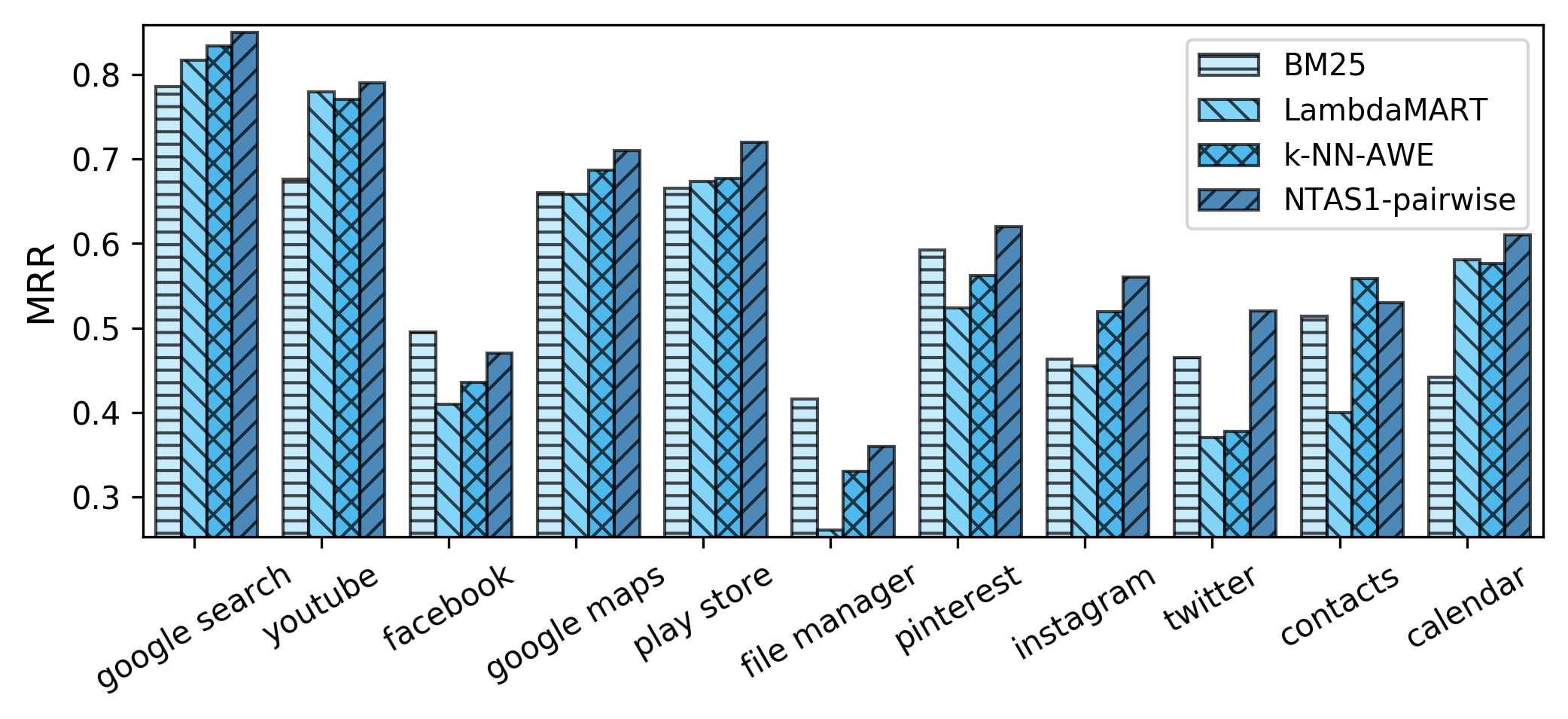}
    }
    \caption{Performance comparison with respect to certain apps on both data splits.}
    \label{fig:res_app}
\end{figure}

\partitle{Performance on apps.}
Here, we compute the mean performance of the queries targeted to a specific app and plot the result for each app in Figure~\ref{fig:res_app}. For the sake of visualization, we only compare the performance of \ntar1-pairwise with three other methods in terms of MRR. We see that all models perform well in ranking less personal apps such as Google Search, YouTube, and Google Maps. Since none of the models incorporate users' personal data, this result is expected. This suggests users' personal data can be leveraged to rank apps such as File Manager, Contacts, and Calendar higher. 
Also, users' activities on their social media apps should be leveraged to provide a more effective personalized ranking.
% Also, one's social media data should be used to rank social media apps. 
Moreover, we see that k-NN-AWE is more robust across the two data splits, compared to other baselines. In particular, it performs well in ranking Contacts suggesting that the proximity of contact names in the high-dimensional space of word embedding enables k-NN-AWE to outperform other models.
Finally, it can be seen that \ntar1-pairwise is more robust across the two data splits, compared to other methods. Specifically, it outperforms all other methods for the majority of apps. However, \ntar1-pairwise performs worse than other methods for File Manager, on both data splits. 
This is mainly due to insufficient number of training data, given the diversity of the queries related to this app.
% The reason is that although the queries for this app are highly diverse, the number of training data is limited.
%might be related to the limited number of queries available for this app in \mohammad{can you think of any reason?}

\begin{figure}
    \centering
    \subfloat[UniMobile-Q Dataset]{
        \includegraphics[width=.475\columnwidth,valign=t]{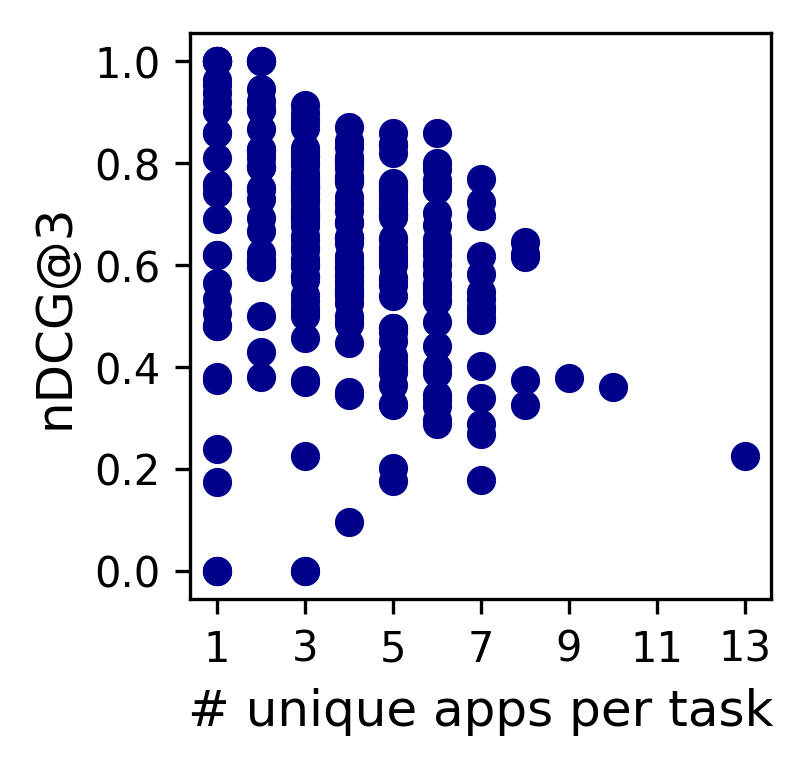}
    }
    \subfloat[UniMobile-T Dataset]{
        \includegraphics[width=.475\columnwidth,valign=t]{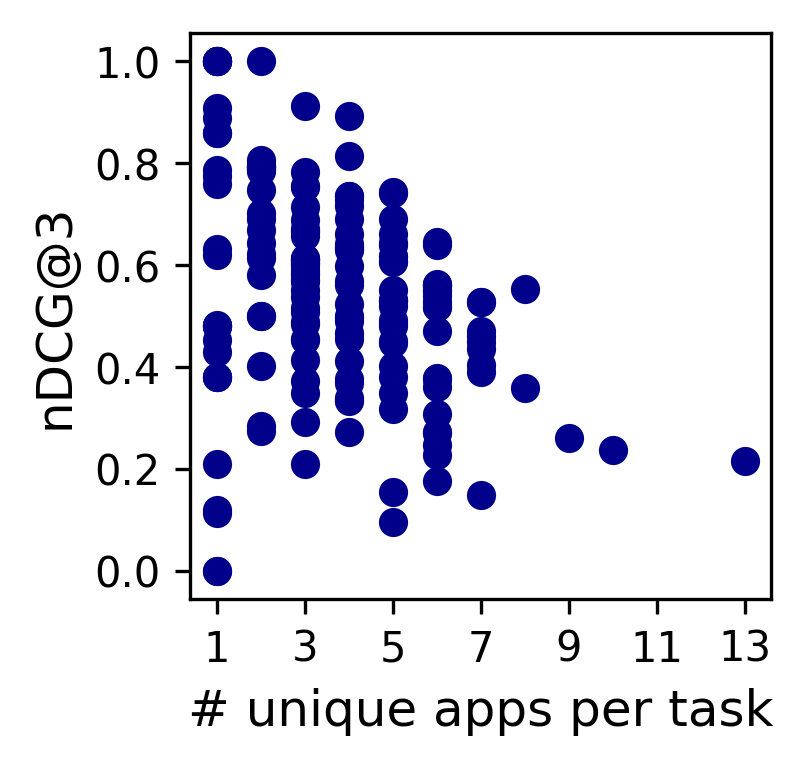}
    }
    \caption{Negative correlation between the number of unique apps users selected for a task and performance.}
    \label{fig:res_task}
\end{figure}

\partitle{Performance on tasks.}
We are interested in seeing how methods perform differently with respect to different search tasks. To do so, we averaged the performance (nDCG@3) of all queries belonging to the same task. Then, we grouped the tasks by the total number of unique apps selected by users and plotted their results in Figure~\ref{fig:res_task}. Our intuition was that if different users chose several apps for a single task, it can be a sign that the task is more challenging for the models. We can see in the figure that as the number of unique apps per task raises, the models perform worse. Although, the negative correlation is not very strong (Pearson: $-0.3049$ and $-0.3450$ for \unimobile-Q and \unimobile-T, respectively), it is consistent with all models and evaluation metrics. This indicates that if a task can be done using multiple apps, their corresponding queries also become more difficult for a system. A multi-app task can be either very personal (i.e., every user chooses their own favorite app) or very general (i.e., it can be done using many apps). Therefore, one can explore incorporating users' regular app usage patterns to perform a personalized target app selection.

\section{Conclusions and Future Work}
In this paper, we introduced and studied the task of target apps selection, which was motivated by the growing interest in conversational search systems where users speak their queries to a unified voice-based search system.
To this aim, we presented the first analysis of mobile cross-app search queries and user behaviors in terms of the apps they chose to complete different search tasks. 
We found that a limited number of popular apps attract most of the search queries. We further observed notable differences between queries submitted to different apps. We showed that query length and content differ among apps. We also showed that, 39\% of search queries were done in Google Search, and it was the top choice of users in 35\% of the tasks. Given that more than 71\% of the defined tasks could be done with the current features of Google Search, this indicates that users prefer to search using a more specific app. 
We carried out the experiments and analyses on the dataset of cross-app mobile queries that we collected through crowdsourcing.

Since the mobile information environment is uncooperative and the data is heterogeneous, representing each app for the target apps selection task is challenging. We proposed two models that learn high-dimensional latent representations for the mobile apps in an end-to-end training setting. Our first model produces a score for a given query-app pair, while the second model produces a probability distribution over all the apps given a query.
We compared the performance of our proposed method with state-of-the-art retrieval baselines splitting data following two different strategies. Our approach outperformed all baselines significantly. 

There are several directions for future work. We plan to conduct a follow-up study asking volunteers to install an app which will track their movement and sense their context. We will ask the volunteers to report their daily mobile search experiences using our app. This will enable us to study user behaviors while searching with different apps in the wild. Since we will not ask users to complete predefined search tasks, we expect to see different distribution of search tasks and selected apps. Moreover, a real unified mobile search system would have access not only to the users' personal selection of apps, but also to their daily app usage patterns. Incorporating such information into the ranking model is an interesting future direction. More importantly, mobile devices can be used to sense users' context. Another future direction is to study how the sensed contextual information can be leveraged to enhance a ranking model. Also, search results aggregation and presentation should be explored in the future, considering two important factors: high information gain and user satisfaction.

% Further future directions are related to results presentation that was not explored in this work. Results aggregation and presentation should be explored in the future, considering two important factors: high information gain and user satisfaction. This direction can be explored in both areas of information retrieval and human-computer interaction. It would be interesting to see if users prefer to see an aggregated list of results (similar to works done in federated search), or if they prefer to see results grouped into their source app. In addition, based on our findings in the analyses, we believe that mobile search queries can be leveraged to improve the user experience. For instance, a user looks for a restaurant using a unified search system and finds the relevant information in Yelp. Considering the users' personal preference as well as their context, the system can push a notification about the traffic near the restaurant as soon as the user starts driving towards it. 

\medskip

\begin{acks}
% {\footnotesize
% \noindent \textbf{Acknowledgements.} 
This work was supported in part by the RelMobIR project of the \grantsponsor{}{Swiss National Science Foundation (SNSF)}{http://www.snf.ch/en/Pages/default.aspx}, in part by the Center for Intelligent Information Retrieval, and in part by NSF grant \#IIS-1160894. Any opinions, findings and conclusions or recommendations expressed in this material are those of the authors and do not necessarily reflect
those of the sponsors.
% }

% This research was partially funded by the RelMobIR project of the \grantsponsor{}{Swiss National Science Foundation (SNSF)}{http://www.snf.ch/en/Pages/default.aspx}.
\end{acks}

\bibliographystyle{ACM-Reference-Format}
\bibliography{sigproc} 

\end{document}